\documentclass[twocolumn,english]{revtex4-2}
\usepackage[utf8]{inputenc}
\usepackage[T1]{fontenc}
\usepackage{color}
\definecolor{document_fontcolor}{rgb}{0, 0, 0}
\color{document_fontcolor}
\usepackage{babel}
\usepackage{textcomp}
\usepackage{amsmath}
\usepackage{graphicx}
\usepackage[unicode=true,pdfusetitle,
 bookmarks=true,bookmarksnumbered=false,bookmarksopen=false,
 breaklinks=false,pdfborder={0 0 1},backref=false,colorlinks=false,pdfpagemode=FullScreen]
 {hyperref}

\makeatletter
\numberwithin{equation}{section}
\numberwithin{figure}{section}

\makeatother

\begin{document}
\title{\textcolor{black}{Characterization of the optical response from variant
InGaN nanowires emitting within the green spectral gap}}
\author{\textcolor{black}{M. Esmaeilzadeh}\textsuperscript{\textcolor{black}{1,2}}\textcolor{black}{,
P. Tieben}\textsuperscript{\textcolor{black}{1,2}}\textcolor{black}{,
S. Chatterjee}\textsuperscript{\textcolor{black}{3}}\textcolor{black}{,
A. Laha}\textsuperscript{\textcolor{black}{3}}\textcolor{black}{,
A. W. Schell}\textsuperscript{\textcolor{black}{1,2,4}}}
\affiliation{\textcolor{black}{1. Physikalisch-Technische Bundesanstalt, Bundesallee
100, 38116 Braunschweig, Germany}}
\email{m.esmaeilzadeh1990@gmail.com}

\affiliation{\textcolor{black}{2. Institut for Festk\"{o}rperphysik, Leibniz Universit\"{a}t
Hannover, Appelstr. 2, 30167 Hannover, Germany}}
\affiliation{\textcolor{black}{3. Department of Electrical Engineering, Indian
Institute of Technology Bombay,400076 Mumbai, India}}
\affiliation{4. Institute of Semiconductor and Solid State Physics, Johannes Kepler
University Linz, Altenberger Stra\ss e 69, 4040 Linz, Austria}
\begin{abstract}
This study provides a comprehensive physical and optical investigation
of InGaN nanowires (NWs) designed to address the challenges posed
by the green gap region. We conduct a detailed analysis of the morphology,
structure, and optical characteristics of the NWs using \textcolor{black}{characterization}
techniques such as scanning electron microscopy, cathodoluminescence
spectroscopy, and confocal scanning microscopy. Notably, increasing
the indium concentration causes a redshift in emission and alters
the luminescence properties across different segments of NWs. 

Our findings provide valuable insight into the correlation between
indium compositional nonuniformity and the optical emission properties
of NWs. These insights contribute to optimizing the growth condition,
color accuracy, and enhancing optical efficiency of NWs, highlighting
their potential for next generation high-performance LEDs and optoelectronics
devices.
\end{abstract}
\maketitle

\section{Introduction}

InGaN-based light-emitting diodes (LEDs) are in high demand for next
generation of laser diodes, high-performance LEDs, display technologies,
and other optoelectronic devices due to their ability to emit light
across the entire visible spectrum, particularly in the green gap
region \textcolor{black}{\cite{tsao2014toward, goodman2011green, krishna2013carrier, giftthaler2024green, cheng2020nanoscale, li2018very, park2011green, hahn2011epitaxial, bai2022microscale, liu2016shockley, liu2021high, liu2022n, zhang2016single, zhao2018iii, dupre2017processing, liu2022micro, li2020high}.}

\textcolor{black}{By varying the indium concentration in the active
region during growth of NWs, the emission wavelength can be tuned.
Higher indium content leads to a redshift in the emission spectrum
allowing precise control over the emission color. However, achieving
both high efficiency and good crystal quality at higher indium concentration
remains a significant challenge\cite{banerjee2016superluminescent,chang2010high, fu2018explaining, kuykendall2007complete, murotani2013effects, philip2017high, navid2022gan}. }

\textcolor{black}{In our work, we study InGaN/GaN nanowires (NWs)
fabricated by plasma-assisted molecular beam epitaxy in order to explore
their potential as highly efficient nanostructures emitting in the
green gap region. We analyze the morphology and optical structure
of the NWs using characterization techniques including scanning electron
microscopy (SEM), cathodoluminescence spectroscopy (CL)}, and our
custom-built confocal microscope. The confocal setup enables detailed
mapping of wavelength variations along individual NWs through the
method of spatial spectral scanning, hereafter referred to as Confocal-SSS.
Our findings reveal the effects of spatial variations in indium concentration
on emission properties and compositional homogeneity of the NWs. 

The results of this study may contribute to the development of next
generation InGaN-based nanostructures with improved efficiency, color
accuracy, and spectral purity, as well as better integration with
emerging technologies. 

\section{Results and discussion}

\begin{figure}[h]
\textcolor{black}{\includegraphics[scale=0.3]{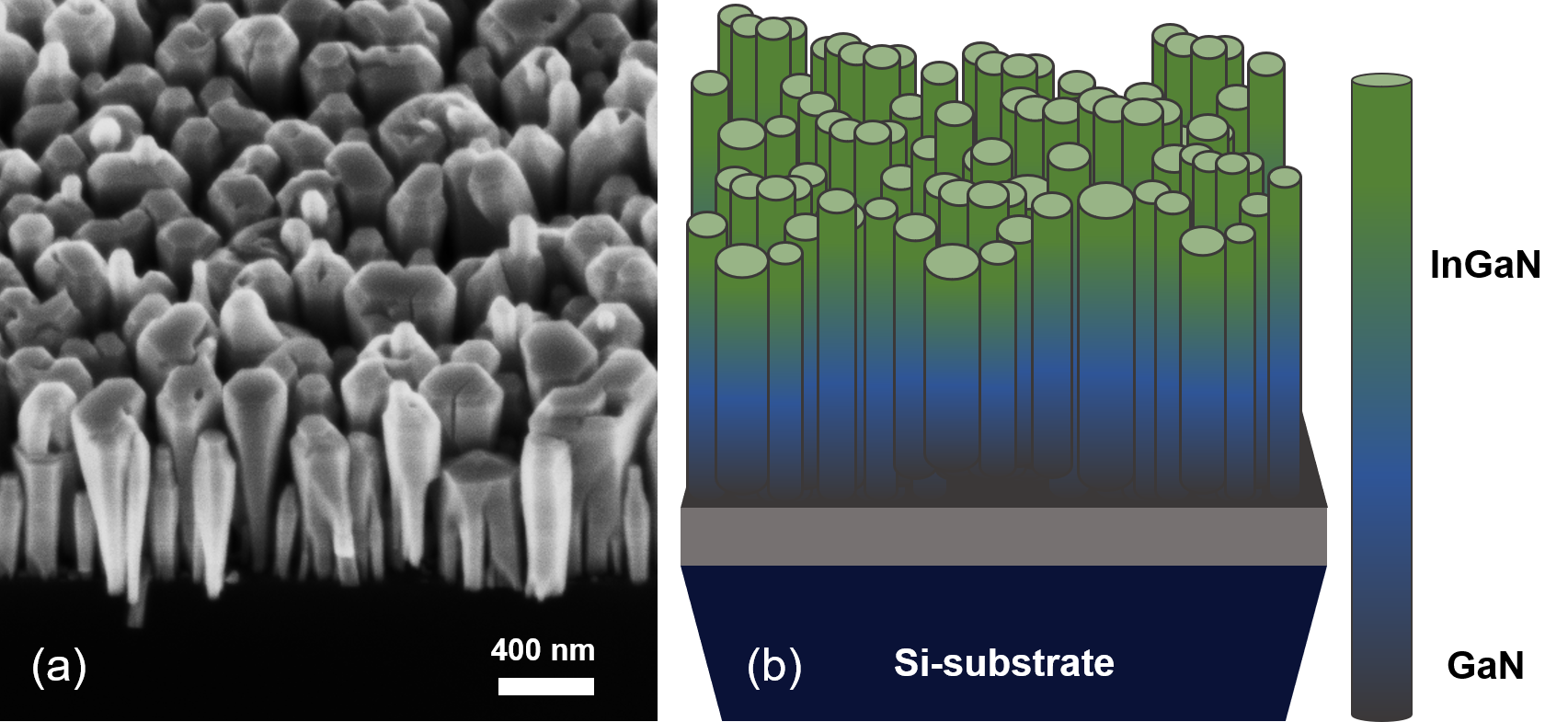}}

\textcolor{black}{\caption{\textcolor{black}{(a) SEM image and (b) schematic representation of
NWs grown on Si substrate composed of a GaN seed layer and InGaN layers.}\textcolor{green}{{}
\label{NW_Schematic }}}
}
\end{figure}

We employ various measurement and analysis techniques to investigate
the morphology, structure, and optical properties of\textcolor{black}{{}
InGaN nanowires (NWs) grown on} Si (111)\textcolor{black}{{} substrate
(see the Experimental Section). Fig. $\text{\ref{NW_Schematic }}$
illustrates} an SEM (Raith, Pioneer II)\textcolor{black}{{} image and
schematic representation of the grown NWs composed of GaN seed layer
and InGaN layers.}

\textcolor{black}{}

\begin{figure}
\textcolor{black}{\includegraphics[scale=0.37]{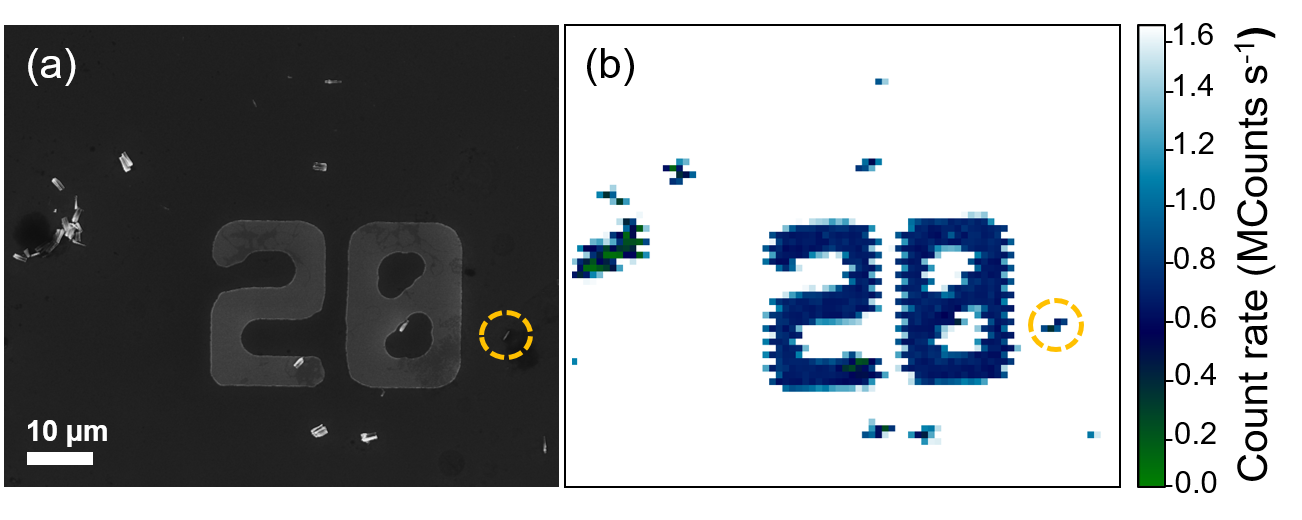}}

\textcolor{black}{\caption{Precise spatial localization of NWs across both SEM and confocal scan
images. The dashed yellow circle highlights our ability for exact
mapping of a NW on the substrate. \label{Locating}}
}
\end{figure}

To investigate the morphological and optical characteristics of the
grown NWs, it is necessary to avoid NW clusters and study each NW
individually. For this, a solution of scraped NWs was prepared and
applied to the measurement substrate via drop-casting method. We also
employed lithography techniques to fabricate mark-imprinted and custom-designed
substrates to precisely locate the positions of individual NWs on
the measurement substrates. \textcolor{black}{Fig. $\text{\ref{Locating}}$}
shows SEM and a confocal scan images of the same NW, illustrating
the effectiveness of the alignment method. The different random orientations
on the substrate is due to the nature of the drop-casting mechanism.

\textcolor{black}{}

\begin{figure}
\textcolor{black}{\includegraphics[scale=0.67]{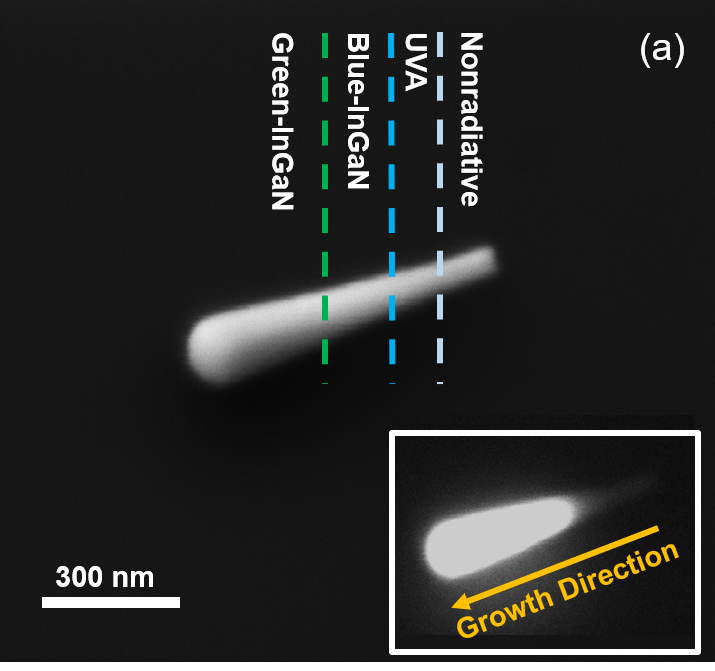}}

\textcolor{black}{\includegraphics[scale=0.35]{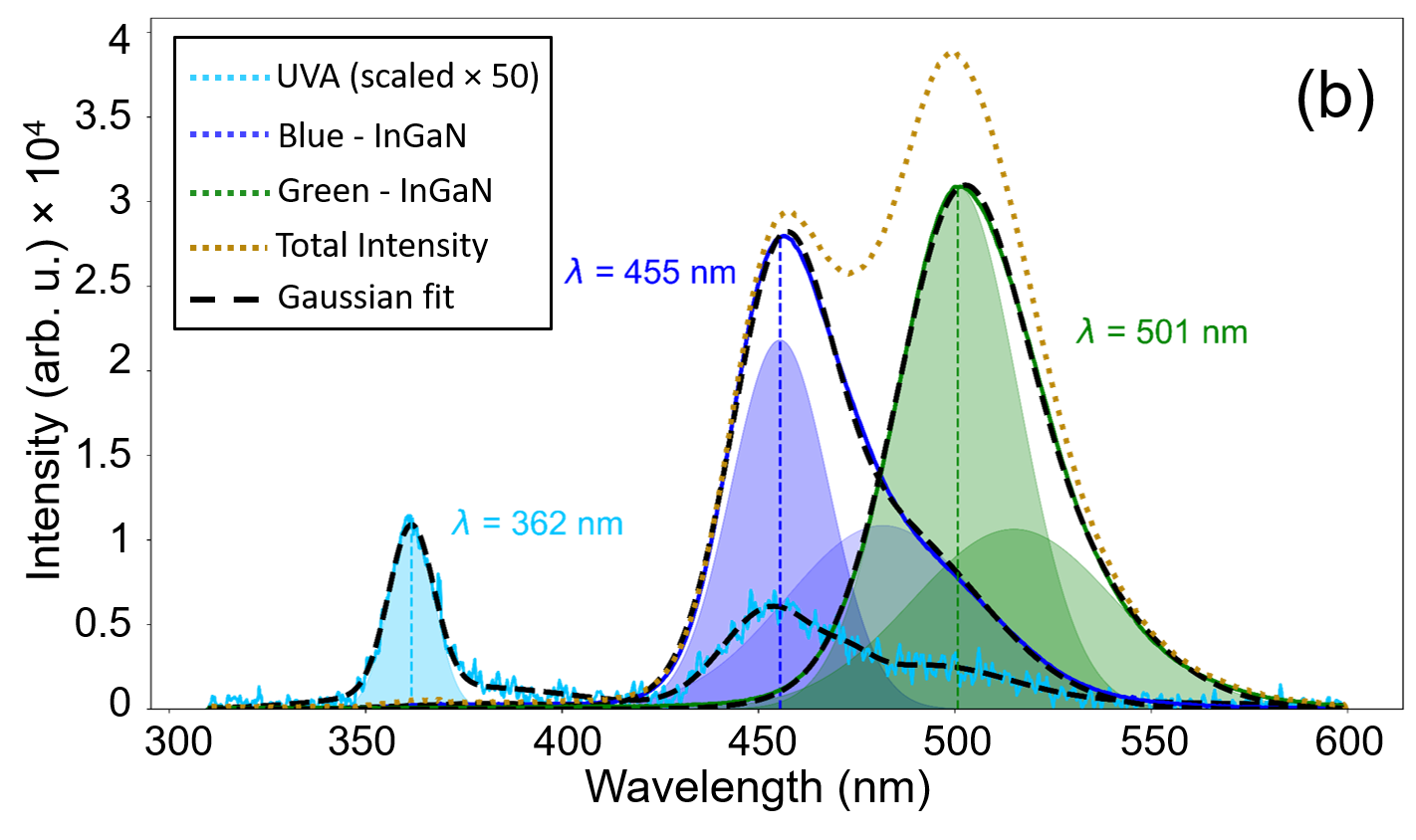}}

\textcolor{black}{\caption{CL analysis of an exemplary NW. (a) represents categorized segments
of the NW based on dominant emission wavelengths. The inset shows
the gradient in emission intensity. (b) shows the CL spectra of different
radiative segments of the NW including, UVA-GaN, Blue-InGaN, and Green-InGaN.
\label{NW_segments}}
}
\end{figure}

To examine the compositional structure of the NWs, we analyzed several
different NWs using CL (TESCAN, MIRA3) technique.\textcolor{black}{{}
Fig. $\text{\ref{NW_segments}}$} shows the CL image of an exemplary
NW composed of various segments, including radiative and non-radiative
parts. The inset of the image exhibits different photoluminescence
levels among different segments of the NW in the visible range. The
presence of the non-radiative part is attributed to the trapped charge
carriers within the structural imperfections during the growth process\textcolor{black}{{}
\cite{hauswald2014origin, toschi2025vn}.} 

\textcolor{black}{Fig. $\text{\ref{NW_segments}}$(b) }shows the corresponding
spectra of the NW represented in Fig\textcolor{black}{. $\text{\ref{NW_segments}}$(a)}
for different segments. The notable redshift in the peak wavelengths
of the spectra along the growth direction of the NW indicates an increase
in indium concentration in the InGaN crystal as reported in other
studies\textcolor{black}{{} \cite{fu2018explaining, kuykendall2007complete}.}

To facilitate clear identification in detailed analysis of radiative
segments, we categorize them as UVA-GaN, Blue-InGaN, and Green-InGaN
segments based on the dominant wavelengths in the emission spectra.
As illustrated in \textcolor{black}{Fig. $\text{\ref{NW_segments}}$,}
the UVA-GaN segment, associated with the GaN seed layer, emits across
a broad spectral range with a peak at 362 nm that is consistent with
the characteristic emission wavelength of GaN\textcolor{black}{{} \cite{nogues2014cathodoluminescence, zagonel2012visualizing}.
Compared to the Blue-InGaN segment, the UVA-GaN exhibits notably dimmer
luminescence, while the Green-GaN segment shows the highest brightness
and efficiency. Moreover, due to varying indium composition within
the InGaN structure, a notable redshift was observed in the emission
spectra of both the Blue-InGaN and the Green-InGaN segments. }

\textcolor{black}{}

\begin{figure}
\textcolor{black}{\includegraphics[scale=0.38]{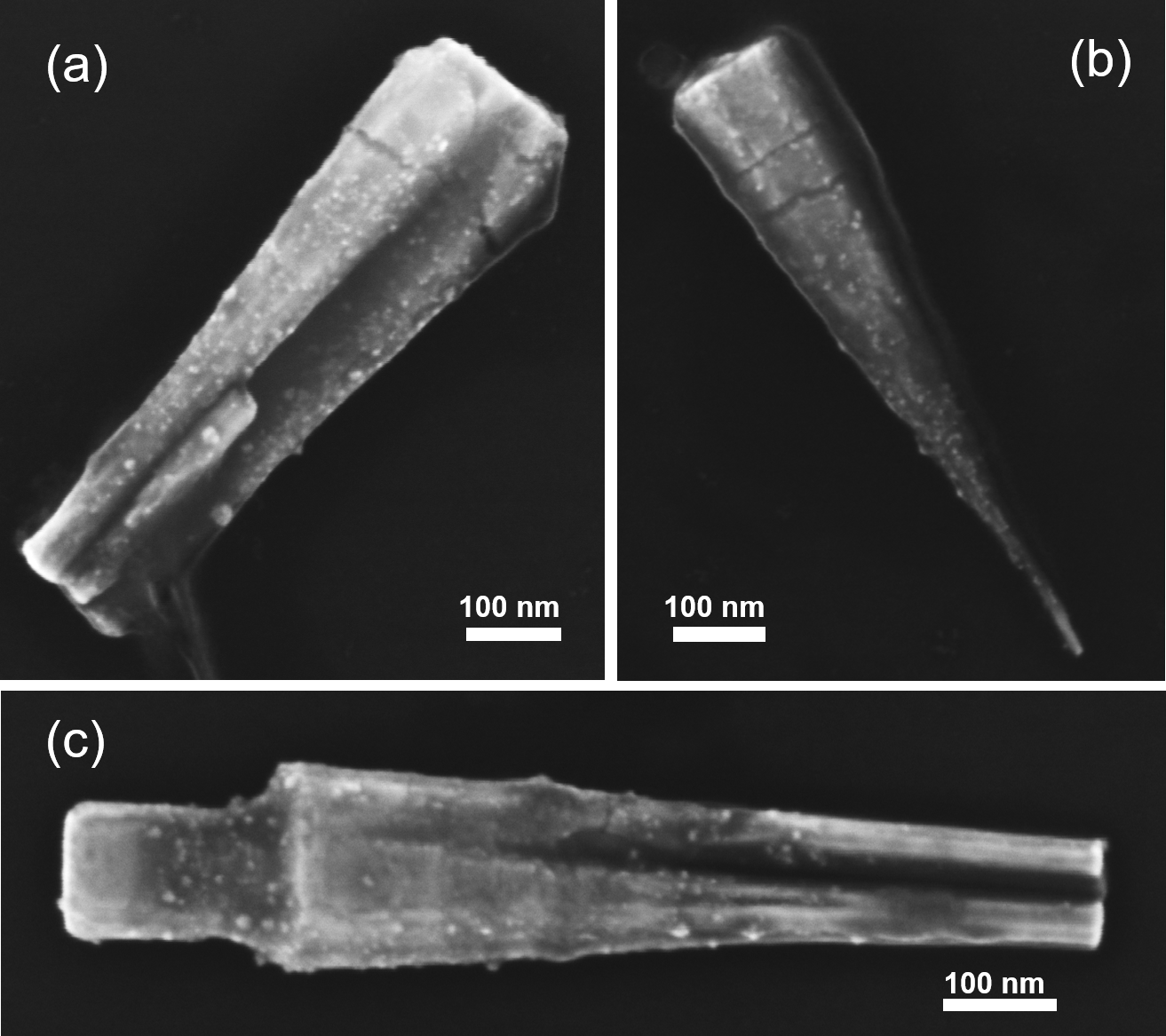}}

\textcolor{black}{\caption{high-resolution SEM image of three randomly selected NWs indicating
different morphologies of NWs. \label{Group_3_NWs_SEM}}
}
\end{figure}

To analyze the morphology of the NWs, we use high-resolution SEM imaging
technique. \textcolor{black}{Fig.$\text{\ref{Group_3_NWs_SEM}}$}
shows SEM images of three exemplary NWs randomly selected with different
sizes and morphologies. According to our analysis, the NWs can have
different lengths ranging from 600-900 nm, different maximum widths
ranging 80 nm to150 nm along the whole body. 

\textcolor{black}{}

By employing Confocal-SSS measurement technique, we are able to study
the optical properties of Blue-InGaN and Green-InGaN segments of NWs
such as photoluminescence brightness level and visible emission spectrum\textcolor{black}{{}
(see the Experimental Section).}

\begin{figure*}
\textcolor{black}{\includegraphics[scale=0.55]{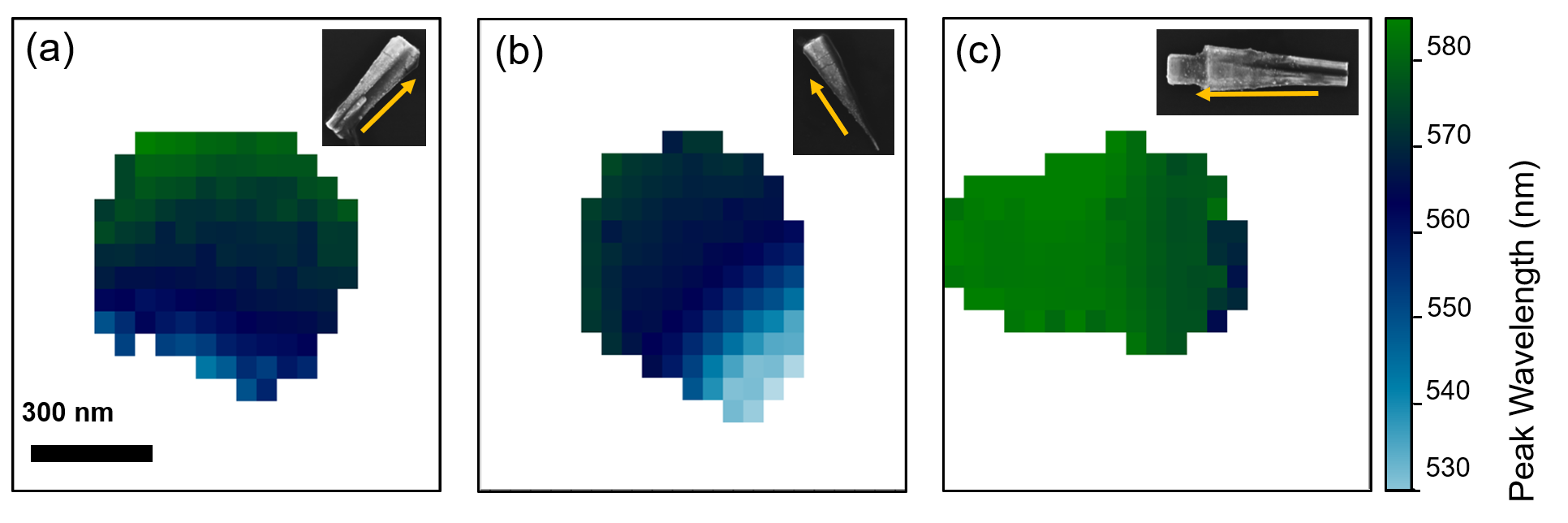} }

\textcolor{black}{\caption{Mapping of peak emission wavelengths of Blue-InGaN and Green-InGaN
segments of the three exemplary NWs presented in \textcolor{black}{Fig.
$\text{\ref{Group_3_NWs_SEM}}$} obtained by Confocal-SSS measurement.
The inset shows the original image of the NWs, with the yellow arrow
indicating their growth direction. The observed redshift highlights
compositional inhomogeneity along the NW axis.The redshift highlights
compositional inhomogeneity along NWs.\label{3D_spectra-1}}
}
\end{figure*}

\textcolor{black}{Fig. $\text{\ref{3D_spectra-1}}$}, represents the
peak emission wavelength of exemplary NWs introduced in \textcolor{black}{Fig.
$\text{\ref{Group_3_NWs_SEM}}$} at each scanned pixel derived from
the fitting process. The growth direction of the NWs is clearly identifiable
in the SEM images and remains consistent in the confocal scan, confirming
the accuracy and alignment of the measurements setup. 

As can be seen from \textcolor{black}{Fig. $\text{\ref{3D_spectra-1}}$},
there are significant redshifts along the growth direction of the
NWs exceeding 20, 40, and 80 nm for NW (c), (b), and (a) respectively.
The redshift observed along the NWs indicates a variation in indium
incorporation developed during the growth process. In fact the variation
in spectral shape observed at different positions along a NW arises
from compositional differences, causing inhomogeneity in redshift
magnitude. The dominant luminescence of the NWs ranges from 520 nm
to 580 nm addressing the desired emission wavelengths within the green
gap region. This evidence supports that our fabricated nanostructures
are potential candidates for the next generation of optoelectronics
devices. 

\textcolor{black}{}

\section{Conclusion}

In summary, in this study, we provide a comprehensive physical and
optical analysis of the InGaN nanostructures developed in our group
using different techniques, including SEM, CL, and Confocal-SSS. 

Our findings highlight that compositional nonuniformity in indium
concentration causes a redshift in the spectral emission. This nonuniformity
can lead to variation in luminescence spectrum and intensity across
different segments of the NWs, with the most efficient emission in
the green gap region.

The observed correlation between the indium concentration gradient
and the resulting redshift in emission provides valuable insight for
optimizing the growth process of the NWs and enhancing the optical
efficiency and color accuracy of the nanostructures. This will pave
the way for future advances in optoelectronic devices and high-performance
LEDs emitting within the green gap region. 

\section{Experimental Section}

\textbf{MBE Growth:}\textcolor{black}{{} The growth process of NWs was
conducted using plasma-assisted molecular beam epitaxy (RIBER, MBE
C21 system) equipped with a UNI-Bulb plasma cell (Veeco) \cite{bhunia2023scaling, nag2020role}}.
The process begins with the nucleation of a GaN seed layer on the
substrate, forming the initial GaN nanocrystal nuclei. This is followed
by epitaxial growth of InGaN NWs on top of the seeds\textcolor{black}{{}
\cite{chatterjee2024role}}.

In the first step, the RCA-cleaned and degassed Si (111) samples were
annealed in the growth chamber at 860 \textdegree C until surface
reconstruction was observed. Then the surface was nitridated for 15
minutes at 800 \textdegree C, enabling Volmer-Weber growth by modifying
the surface energy. Then, GaN was grown for 45 minutes under N-rich
conditions, forming 3D nuclei, acting as a base for NW growth\textcolor{black}{{}
\cite{sarkar2016comprehensive}}. These nuclei are usually 50-100
nm in height.

To proceed with the growth process of the InGaN, the temperature was
reduced to 640 \textdegree C facilitating indium incorporation since
indium atoms have higher mobility and vapor pressure compared to gallium
atoms.

However, reducing growth temperature also results in widening of the
NW tip due to enhanced lateral growth. This occurs because the reduced
surface diffusion length of indium atoms limits their migration toward
the NW base, causing more deposition near the tip and consequently
widening it\textcolor{black}{{} \cite{pan2023desorption, tang2015effect}.}
Now the InGaN grown on the nucleated GaN base for about $3$ hours
with an indium flux of $4.0\times10^{-7}$ Torr and gallium flux of
$1.6\times10^{-7}$ Torr. The grown NWs have lengths ranging from
600-900 nm. By maintaining nitrogen plasma at 500 W and a flow rate
of 3.5 sccm, we can control the diameter of NWs and ensure ideal nitrogen-rich
conditions for III-N NWs growth\textcolor{black}{{} \cite{fernandez2009growth}}. 

\textbf{Confocal-SSS setup: }To study optical properties of NWs targeted
to emit within the green spectrum, we used confocal microscopy technique
to scan the NWs. The Confocal-SSS setup is designed to cover a spectral
range from 430-600 nm using a laser source emitting at 405 nm for
excitation and an air objective (Olympus; MPLAPON) with numerical
aperture of 0.95 to focus the laser light on the sample. The fluorescence
emission intensity was measured with avalanche photodiodes (Laser
Components, COUNT-100C-FC) using a $20\times20$ pixel spatial scan
over a 1\textmu m\texttwosuperior{} area. For spectral analysis of
NW\textquoteright s emission, we used a spectrograph (Princeton Instruments,
SpectraPro HRS500) equipped with a CCD camera (Princeton Instruments,
PIXIS: 100B). All spectra were recorded with a grating constant of
150 lines/mm and an integration time of 200 ms. 

\section*{Acknowledgements}

The authors gratefully acknowledge financial support from the Alexander
von Humboldt Foundation through the reasearch travel grant AvH Ref
3.5 - 1118355 - IND - HFS awarded to A. Laha.

\section*{keywords }

light emitting devices, nanostructures, indium gallium nitride, cathodoluminescence,
scanning electron microscopy, photoluminescence

\bibliographystyle{plain}
\bibliography{sorsamp}
\end{document}